\documentclass[sigconf]{acmart}
\usepackage{graphicx}
\usepackage[labelfont=bf]{caption}
\usepackage{color}

\usepackage{todonotes}

\usepackage{amsmath,amssymb,amsfonts,mathrsfs}
\usepackage{booktabs}
\usepackage{caption,subcaption}

\copyrightyear{2019} 
\acmYear{2019} 
\acmConference[ICTIR '19]{The 2019 ACM SIGIR International Conference on the Theory of Information Retrieval}{October 2--5, 2019}{Santa Clara, CA, USA}
\acmBooktitle{The 2019 ACM SIGIR International Conference on the Theory of Information Retrieval (ICTIR '19), October 2--5, 2019, Santa Clara, CA, USA}
\acmPrice{15.00}
\acmDOI{10.1145/3341981.3344239}
\acmISBN{978-1-4503-6881-0/19/10}

\begin{document}

\title[DC$^3$ -- A Diagnostic Case Challenge Collection]{DC$^3$ -- A Diagnostic Case Challenge Collection for\\Clinical Decision Support}

\author{Carsten Eickhoff}
\affiliation{%
  \institution{Brown University}
  \institution{codiag AG} 
  \country{USA} 
}

\author{Floran Gmehlin}
\author{Anu V.\ Patel}
\author{Jocelyn Boullier}
\affiliation{%
 \institution{codiag AG}
 \country{Switzerland} 
}

\author{Hamish Fraser}
\affiliation{%
	\institution{Brown University}
	\country{USA} 
}

\begin{abstract}
In clinical care, obtaining a correct diagnosis is the first step towards successful treatment and, ultimately, recovery. Depending on the complexity of the case, the diagnostic phase can be lengthy and ridden with errors and delays. Such errors have a high likelihood to cause patients severe harm or even lead to their death and are estimated to cost the U.S. healthcare system several hundred billion dollars each year. 

To avoid diagnostic errors, physicians increasingly rely on diagnostic decision support systems drawing from heuristics, historic cases, textbooks, clinical guidelines and scholarly biomedical literature. The evaluation of such systems, however, is often conducted in an ad-hoc fashion, using non-transparent methodology, and proprietary data. 

This paper presents DC$^3$, a collection of 31 extremely difficult diagnostic case challenges, manually compiled and solved by clinical experts. For each case, we present a number of temporally ordered physician-generated observations alongside the eventually confirmed true diagnosis. We additionally provide inferred dense relevance judgments for these cases among the PubMed collection of 27 million scholarly biomedical articles.
\end{abstract}

%
%



\maketitle

\renewcommand{\shortauthors}{C.\ Eickhoff, F.\ Gmehlin, A.\ V.\ Patel, J.\ Boullier and H.\ Fraser}

\section{Introduction}\label{sec:intro}

Diagnostic errors refer to the failure to establish an accurate and timely explanation of the patient's health problem(s) or to communicate that explanation to the patient~\cite{national2016improving}. With relative shares of up to 40\%, several studies report diagnostic errors to constitute the largest and most impactful source of avoidable primary care error~\cite{phillips2004learning,sandars2003frequency,singh2010reducing}, leading to the most considerable share of claims~\cite{silk2000went,gandhi2006missed,singh2013types} against primary care physicians. In the USA, an annual economic damage of hundreds of billions (a sizable portion of the country's overall health spendings) is attributed to diagnostic errors~\cite{mcginnis2013best}. Given the key importance of correctness and timeliness of primary care diagnosis, a large body of work has been investigating systematic reasons for misdiagnoses in this setting. Numerous studies list low disease prevalence as one of the top causes of diagnostic errors and delays in primary care~\cite{kostopoulou2008diagnostic,sarkar2012challenges,singh2013types}, as uncommon diagnoses may be overshadowed by more prevalent ones in the cognitive diagnostic process~\cite{jopp2001diagnostic,sarkar2012challenges}. In an effort to improve patient safety, there are frequent calls for more effective diagnostic processes in primary care~\cite{lorincz2011research,wynia2011improving}, involving a greater utilization of electronic health record (EHR) and clinical decision support systems~\cite{newman2009diagnostic}. Complex patients with non-specific presentations, multiple co-morbidities, and rare conditions are assumed to be at an especially high risk of receiving a delayed or inaccurate diagnosis~\cite{national2016improving}.

Clinical decision support systems aim to help health professionals in addressing particularly challenging diagnosis or treatment needs~\cite{galko2018biomedical,grnarova2016neural,meyer2018machine,wei2018embedding}. In order to assess the accuracy of such systems, annotated examples of patient case information are required. To date, there are only few such resources available to researchers. As a consequence, many clinical decision support system evaluation campaigns rely on small, outdated or proprietary sources of data, making their findings difficult to verify and reproduce.

This paper presents DC$^3$, a collection of 31 extremely difficult diagnostic case challenges, that were manually compiled and solved by clinical experts. For each case, there are a number of temporally ordered physician-generated observations alongside the eventually confirmed true diagnosis. We additionally provide inferred dense relevance judgments for these cases among the PubMed collection of 27 million scholarly biomedical articles.

\section{Comparison to Existing Collections}
The vast majority of clinical decision support systems are trained and evaluated on proprietary samples of patient data that, for reasons of confidentiality, cannot be released to the research community. There are, however, a number of openly available collections that deserve mentioning. 

\textbf{TREC CDS.} The TREC Clinical Decision Support (CDS) track~\cite{simpson2014overview} was run from 2014 to 2016 and tasked participating systems to retrieve biomedical literature in response to 90 natural-language patient descriptions. The patients stem from general intensive care populations with relatively common indications. For the majority of these patients no explicit target diagnosis information is available. All patient descriptions are single-shot narratives in the style of an anamnesis without temporally ordered findings.

\textbf{CLEF eHealth.} Similar to the TREC CDS track, the 2013 to 2015 editions of CLEF's eHealth challenge~\cite{suominen2013overview} offered a patient-centric information retrieval task. Given a minimal patient profile, the participants were tasked with retrieving relevant literature related to the case. 

\textbf{MIMIC-III.} The MIMIC critical care database~\cite{johnson2016mimic} contains electronic records of 59,000 intensive care admissions collected between the years of 2001 and 2012. Each admission specifies a list of diagnostic and procedural ICD codes that were designed for billing purposes but can be used as targets for diagnostic decision support. Given the intensive care domain, the range of observed diagnoses is limited in comparison to the complex cases outlined here.

\textbf{i2b2.} The i2b2 initiative has been launching a number of classification and information extraction benchmarking campaigns for which annotated corpora of de-identified patient records were provided. Most studied tasks focus on extracting patient properties such as smoker status~\cite{uzuner2008identifying} or obesity~\cite{uzuner2009recognizing} that are of only limited interest in the diagnostic decision support use case. Others are limited to a single class of diagnoses such as heart disease~\cite{stubbs2015identifying} and do not lend themselves to wide-coverage primary care research.

To the best of our knowledge, at the time of writing this paper, there exists no publicly available dataset of challenging authentic diagnostic episodes that additionally provide both confirmed diagnoses as well as dense query-document relevance judgments for a realistically-sized document collection. DC$^3$ aims to close this gap.

\section{Dataset Details}
The dataset compiles 31 especially challenging diagnostic cases encountered at Massachusetts General Hospital (MGH) in Boston, MA between the years 2013 and 2018.

\subsection{Cases}
Each case is described in a number of topically coherent paragraphs that correspond to the sections usually found in case notes. For the purpose of data extraction, we assume each paragraph to denote an episode in a health record/note entry. Example paragraphs include presenting complaint, history of presenting complaint, examination or investigation and may have been authored by changing physicians (\textit{e.g.}, the first note describes the patient's anamnesis taken by the emergency department's staff while the following note might be written by a radiologist, discussing the findings of a CT scan, \textit{etc}). The cases featured in this corpus are all complex and difficult. While a good proportion of the featured diagnoses are rather uncommon in developed-world hospitals such as MGH (\textit{e.g.}, 16 cases of infectious diseases, or a lead poisoning), they can be more frequently observed in large parts of the developing world. Aside from low-prevalence, many cases have multiple correct diagnoses that jointly account for the patient's symptoms. We represent diagnoses in terms of the corresponding concept unique identifiers (CUI) of the Unified Medical Language System (UMLS). The average case in the collection has 6.9 target CUIs. Finally, we annotate the 2018 snapshot of the National Library of Medicine's PubMed database, identifying all papers whose abstracts mention any of the target diagnoses. In order to achieve this, we rely on a proprietary medical named entity recognition system. Any paper whose title or abstract mention the target diagnoses of a case is considered relevant for that case. The assumption here is that presenting a physician with a paper mentioning the correct diagnosis for the current patient will bring that diagnosis to their attention and thereby increase the chance of them testing for and eventually confirming it. Unlike many existing IR benchmarking collections, that rely on pooled manual relevance assessment, this approach results in dense\footnote{There is the possibility of NER false negatives that would lead to missed potentially relevant documents. Given the generally high performance of this system, we consider this a minor risk to corpus quality.} relevance labels. The median number of relevant documents per case is 3,597. Especially for outlier cases with many target diagnoses, such numbers are much higher than what is observed in typical Web search collections with only partial relevance labels. Table~\ref{tab:data} gives a complete overview of all cases. We did not perform any manual sub-selection of cases and, instead, included all currently published MGH case challenges.

\begin{table*}[]
    \centering
    \caption{DC$^3$ Case Details}
    \label{tab:data}
    \begin{tabular}{lrp{10cm}rr}
        \toprule
        Case ID & \# Notes & Final Diagnosis & \# CUIs & \# rel.\ Docs.\\
        \midrule
        1 & 10  & Histoplasma capsulatum infection &3& 121\\
        2 & 11  & Necrotizing lymphadenitis, with features consistent with histiocytic necrotizing lymphadenitis (Kikuchi-Fujimoto disease) and scattered EBV-positive cells & 3 & 3,918\\
        3 & 7   & Infective endocarditis and infectious aortitis due to Staphylococcus aureus & 18 & 10,767\\
        4 & 8   & Lead poisoning & 1 & 2,387\\
        5 & 13  & Measles & 7 & 13,466\\
        6 & 7   & Wilson's Disease & 1 & 1,298\\
        7 & 11  & Lemierre's syndrome caused by Fusobacterium necrophorum, with cavernous-sinus thrombophlebitis, carotid-artery thromboarteritis, and abscesses of the parotid gland and subperiosteal orbit & 9 & 276\\
        8 & 7   & Acute anaphylaxis due to a hepatic hydatid cyst caused by Echinococcus granulosus. & 10 & 11,638\\
        9 & 6   & Invasive Neisseria meningitidis infection and primary C8 deficiency & 5 & 3,149\\
        10 & 10 & Perforation of the right ventricular wall (by an implantable cardioverter-defibrillator lead) & 2 & 319\\
        11 & 8  & Borrelia miyamotoi infection and possible Borrelia burgdorferi infection & 4 & 11,044\\
        12 & 11 & Disseminated pulmonary blastomycosis involving the hilar lymph nodes and spleen, early hepatic cirrhosis, and acute tubular necrosis & 5 & 55,750\\
        13 & 10 & Disseminated Mycobacterium bovis infection & 6 & 54,359\\
        14 & 14 & Chronic recurrent abdominal pain caused by intermittent torsion of an accessory spleen & 4 & 794\\
        15 & 25 & Tuberculous enteritis & 4 & 6,501\\
        16 & 7  & Mixed-cellularity subtype of classic Hodgkin's lymphoma and Epstein-Barr virus infection & 10 & 7,538\\
        17 & 9  & Secondary syphilis with neurologic, ocular, and otologic involvement & 13 & 1,774\\
        18 & 8  & Milk Alkali Syndrome & 1 & 343\\
        19 & 9  & Acute HEV infection (Acute Viral Hepatitis) & 13 & 9,222\\
        20 & 12 & Granulomatous amebic encephalitis, caused by acanthamoeba species. Sarcoidosis (old, burned out) involving heart, lungs, and spleen. Coronary arteriosclerosis with stent stenosis. Papillary renal-cell carcinoma and benign biliary hamartoma & 29 & 219,540\\
        21 & 8  & Primary adrenal insufficiency (Addison's disease) & 2 & 864\\
        22 & 7  & IgA vasculitis (Henoch Schonlein Purpura - HSP) & 2 & 23,123\\
        23 & 11 & Acute Leptospirosis & 7 & 3,597\\
        24 & 7  & Acute and chronic cholecystitis and extensive cholelithiasis with transmural gallbladder inflammation & 19 & 19,907\\
        25 & 11 & Combined inherited and acquired thrombotic thrombocytopenic purpura & 4 & 3,233\\
        26 & 10 & Congenital rubella syndrome & 6 & 6,545\\
        27 & 8  & Mycobacterial epididymo-orchitis due to Mycobacterium tuberculosis & 10 & 91,855\\
        28 & 6  & Digoxin toxicity & 2 & 387\\
        29 & 8  & Well-differentiated pancreatic neuroendocrine tumor, grade 1 & 3 & 3,420\\
        30 & 6  & Complete androgen insensitivity syndrome & 5 & 1,219\\
        31 & 11  & Lyme meningoradiculitis & 2 & 135\\
        \bottomrule
    \end{tabular}
\end{table*}

The narrative content of the case notes is written by the treating physicians of the original cases and is composed to reflect the temporal order of discoveries, hypotheses and tests performed, but does not directly reveal the target diagnosis.

\subsection{Distribution Format}
\begin{center}
	\begin{figure}
	\includegraphics[width=0.85\columnwidth]{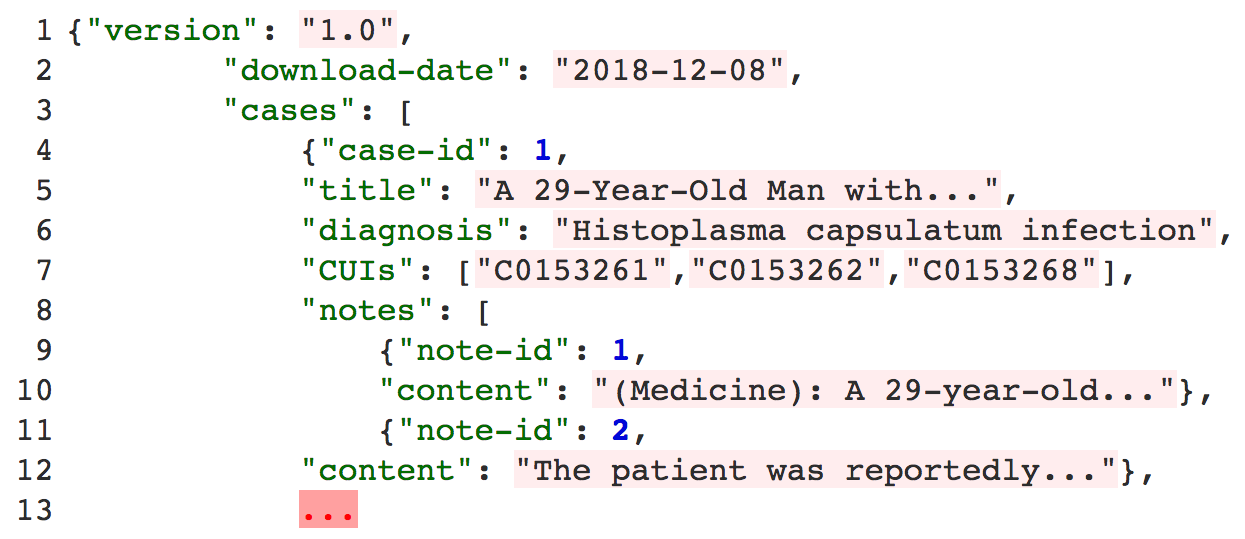}
		\caption{An example of the DC$^3$ case file format.}\label{fig:format}
	\end{figure}
\end{center}

All cases presented in DC$^3$ are originally published in the New England Journal of Medicine's \textit{Case Challenge} section. The copyright remains with them and we do not redistribute any of the case content directly. Instead, we provide the research community with a convenient Python script\footnote{\url{https://github.com/codiag-public/dc3}} that downloads the publicly available case challenges and organizes them in the form of a JSON file. Figure~\ref{fig:format} shows an example of the resulting format. Additionally, we collected inferred dense relevance judgments for the 2018 snapshot of the National Library of Medicine's PubMed database of 27 million scholarly biomedical articles. These relevance judgments are provided in standard \texttt{trec\_eval} format.

\section{Tasks \& Baseline Performance}
In this section, we will describe a range of possible experiments on the DC$^3$ collection. We begin with a patient-centric information retrieval task~\cite{kuhn2016implicit,baruah2018brown} in which we measure the ranking performance of models that take the case description as a query and retrieve scholarly literature articles conducive to making the correct diagnosis. Afterwards, we cast the diagnostic decision support task as a supervised text classification problem in which we model the posterior probability of observing the case description given a diagnosis-specific classifier.

\subsection{Patient-centric Document Retrieval}
This task is similar to the one studied in the TREC Clinical Decision Support (CDS) track~\cite{simpson2014overview}. We use Lucene to index all 27M PubMed abstracts and use the full case description as a query. Table~\ref{tab:retrieval} reports ranking performance of a range of well-known retrieval models in terms of nDCG~\cite{jarvelin2002cumulated} scores at this task. 

\begin{table}
	\caption{Patient-centric literature retrieval results.}\label{tab:retrieval}
	\begin{tabular}{lr}
		\toprule
		Model & nDCG\\
		\midrule
		TF-IDF & 0.37 \\
		LambdaMART & 0.41\\
		DRMM & 0.42\\
		\bottomrule
	\end{tabular}
\end{table}

\subsection{Classification}
In an alternative take on the diagnostic decision support problem, we train a range of disease-specific classifiers on the basis of Pubmed abstracts. We collect all Pubmed abstracts mentioning the target diagnosis as training data and assign the class of maximum posterior probability. This classification problem becomes more difficult as increased numbers $k$ of target diagnoses are being considered. We include the top $k = \{500, 1000, 2000\}$ most frequent diagnoses as observed in Pubmed and compare the performance of Na\"{i}ve Bayes, Logistic Regression and Support Vector Machine classifiers. This selection of classification methods is by no means exhaustive and many more modern and sophisticated techniques are expected to perform better. This overview merely demonstrates the complexity of the task and the demand for innovation in order to truly support unconstrained primary care diagnosis. Table~\ref{tab:classification} reports the results of this comparison in terms of $F_1$ scores. In addition to performance differences between models, increased numbers $k$ of considered target diagnoses significantly increase classification difficulty.


\begin{table}
	\caption{Classification results for different choices of $k$.}\label{tab:classification}
	\begin{tabular}{llll}
		\toprule
		Model & 500 & 1000 & 2000\\
		\midrule
		Na\"{i}ve Bayes & 0.13 & 0.08 & 0.04 \\
		Logistic Regression & 0.14 & 0.08 & 0.06 \\
		SVM & 0.17 & 0.09 & 0.07 \\
		\bottomrule
	\end{tabular}
\end{table}

\section{Conclusion}
In this paper, we present the first version of DC$^3$, a diagnostic case challenge collection for evaluation of clinical decision support systems. The corpus compiles 31 challenging cases from Massachusetts General Hospital in Boston, MA alongside their true underlying diagnoses. As an especially interesting property, we share inferred dense relevance judgments for these cases and the 2018 snapshot of the NLM's PubMed database, allowing for robust and reproducible benchmarking of clinical decision support techniques. In an effort to gauge the collection's difficulty, we investigated two common tasks of interest: Patient-centric literature retrieval, and supervised classification for clinical decision support.

This paper describes a piece of early work in progress that has several limitations to be addressed in the future. (1) Purely inferred relevance judgments do not replace manual expert annotations. They do however offer a powerful means of training retrieval systems that aim to bring the true diagnosis to the physician's attention. (2) Concentrating exclusively on complex cases does not reflect the full spectrum of daily diagnostic tasks encountered by physicians but may help address those cases that doctors struggle with most. (3) The classification and retrieval methods presented in this collection paper are not meant to reflect competitive solutions to the problem but rather aim to illustrate task complexity.

\vspace{-5pt}
\bibliographystyle{plain}
\bibliography{ref.bib} 

\end{document}